# Parameter redundancy in Type III functional response models with consumer interference

María Isabel Cabrera Fernández

Laboratorio de Modelos Ecológicos y Tramas Tróficas, Departamento de Biología, Facultad Experimental de Ciencias, Universidad del Zulia,  Maracaibo 4001, Venezuela

gaiagua@gmail.com

**Abstract**

The consumption rate is a process critically important for the stability of consumer–resource systems and the persistence, sustainability and biodiversity of complex food webs. Its mathematical description in the form of functional response equations is a key problem for describing all trophic interactions.

Because some of the functional response models used in this study presented redundancy between its parameters two methods were used to check for parameter redundancy:  the Hessian matrix calculation using automatic differentiation (AD) which calculates derivatives numerically, but does not use finite differences and the symbolic method that calculates a derivative matrix and its rank.

In this work, we found that the models that better describe the functional response of a rotifer is consumer dependant even at low consumer densities, but their parameters can not be estimated simultaneously because parameter redundancy. This means that fewer parameters or parameter combination can be estimated than the original number of parameters in the models. Here, the model parameters that incorporate intra-specific competition by interference in the consumer-resource interaction are not identifiable, suggesting that this problem may be more widespread than is generally appreciated in the literature of food webs.

Including knowledge on competitive interactions in current model predictions will be a necessity in the next years for ecology as ecological models are getting more complex and more real. Identifiability of biological parameters in nonlinear ecological models will be an issue to consider.

Key words: *Brachionus rotundiformis*, functional response, depletion, interference, AIC, identifiability, Hessian matrix, symbolic algebra software,

## 1. Introduction

Feeding relationships through food webs to a large degree determine the structure and functioning of all ecosystems on earth and the functional response, in particular, is one of the essential links for describing food web models. Their mathematical forms and parameterization strongly influence the dynamics and stability of ecological systems (Murdoch and Oaten, 1975; Hassell et al., 1977; Skalski and Gilliam, 2001; Gross et al. 2004; McCann et al. 2005; Morozov and Arashkevich, 2008; Cabrera, 2011).

Functional response models, following early pioneering work of Holling (1959a, 1959b), have been widely used to model the rate of consumption of individual consumers with respect to the density of food resources. There are three basic forms of functional response: Type I, Type II and Type III, where the consumer consumption rate increases linearly, hyperbolically or sigmoidally respectively. The consumption rate increases with the density of food, until it reaches a saturation threshold where the increase in density of food does not increase the food consumption rate. Moreover there are two parameters involved in the feeding process, the rate of encounters with prey or attack rate ($a$) and the handling time with prey ($h$). Independently Beddington (1975) and De Angelis et al., (1975) defined the incorporation of mutual interference between consumers (intraspecific competition) as the product of the encounter rate between consumers (*b*) and the time spent per encounter between consumers (*w*) in a manner analogous to the parameters *a* and *h* of Holling, and where *b.w* is equal to *i*, the interference intraspecific competition parameter. Another formulation to include mutual interference (parameter *m*) was proposed by Arditi and Akçakaya (1990). In previous work, Cabrera (2011) found that in a two chamber chemostat, with a high density of the rotifer *Brachionus rotundiformis*, there was a decrease in the rate of consumption by interference competition between consumers once the growth has reached a steady state. These results are in agreement with those found in *Daphnia* and other zooplankton species (Peters and Downing, 1984; Seitz, 1984; Folt, 1986; Helgen, 1987; Goser and Ratte, 1994; Burns, 1995; Cleuvers *et al*., 1997) where the consumption rate decreases with the increase in density of individuals. In this work following the results obtained by Cabrera (2011), we proceeded experimentally to determine the functional response of *B. rotundiformis* and the effect of the consumer density on the rate of consumption, and specifically to determine whether the effect of interference competition is also important at low consumer densities. However, despite its simplicity, the two fitted models that incorporate mutual interference competition between the consumers, proved to be intrinsically redundant in their parameters, (i.e. fewer parameters or parameter combinations can be estimated than the original number of parameters in the model) (Reich, 1981; Gimenez et al., 2003; Hunter and Caswell, 2009). Notably the encounter rate (or attack rate) and the mutual interference can not be estimated simultaneously leading to a wrong Akaike Information Criterion[1] (AIC) ranking of the models. Taking into account the parameter redundancy, and fitting models which either include or exclude consumer dependence, we conclude that the rate of consumption in this experiment is consumer dependent even at relatively low consumer densities.

---

[1] An estimate of the expected Kullback–Leibler information lost by using a model to approximate the process that generated observed data (full reality). AIC has two components: negative log likelihood, which measures lack of model fit to the observed data, and a bias correction factor, which increases as a function of the number of model parameters.

## 2. Materials and methods

The rotifer *Brachionus rotundiformis* (lorica length: 72.5 to 235 μm ; lorica width 52.5 to 162.5 μm, n = 386) was isolated from a temporal saline lake in the coastal zone of the Maracaibo Estuary and the alga *Chlorella sorokiniana* (diameter: 2.5 to 6.25 μm, n = 123) was isolated from a dam in the area. Both organisms were kept in semi-continuous culture in a Percival environmental cabinet to 6 ‰ salinity, 28 °C and 2000 lux at 12 hours per day. The rotifer was grown in containers of 5 L, with filtered water enriched with vitamin B12 and fed with *C. sorokiniana* every 2-3 days. The medium was renewed every 15 days. The alga was cultivated in 1 L Erlenmeyer flasks in the medium of Rodríguez-López (1964). A cell counting method was used for determining clearance rates and consumption of the rotifer *B. rotundiformis*, consisting of recording the observed change in the number of suspended cells counted before and after an appropriate exposure time to feeding of the rotifers (Peters and Downing, 1984). According to the same author the direct counting can provide more information than any other estimate of the rate of consumption and avoids the danger of radiotracers and the ambiguity of particle counters. Five concentrations of *C. sorokiniana* were employed [125,000; 500,000; 1,000,000; 2,500,000 and 5,000,000 cells / ml] and four concentrations of *B. rotundiformis* [1, 5, 25 and 125 ind / ml]. A total of twenty experiments were carried out, each one with three replicates and the control. All experiments were performed in a volume of 50 ml and lasted 30 minutes. After this time 10 % formaldehyde was added to the vials to set both the algae and the rotifer. A Neubauer chamber was used for counting the alga and a Sedgwick-Rafter camera was used to count rotifers. For estimating clearance rates and consumption Peters' equations were used (Peters and Downing, 1984), which considers the growth of the control as well as the variation of both rates above and below the saturation level. At the beginning of the experiments, the rotifers were introduced at random to be representative of all sizes. The latter was subsequently confirmed statistically by measuring the body size of representative samples of the specimens of *Brachionus* used in the experiments. For statistical analysis of the functional response curves, the procedure proposed in Juliano (2001) was followed, and consists of two steps: (1) To determine the shape of the functional response requires performing a logistic regression between the proportion of prey consumed (*Ne / N*) vs number of offered prey (*N*) (2) Estimation of the parameters of the functional response curves using nonlinear regression between the number of prey consumed (*Ne*) (we divided *Ne* between *T* (total duration of the experiment), for the consumption rate per ind.min$^{-1}$) vs number of prey offered (*N*). For logistic regression analysis SPSS v. 17 statistical package was used whilst for the nonlinear regression analysis GraphPad Prism v.5 statistical package and AD Model Builder v. 11.2.0 were used.

By definition the functional response is an instantaneous rate and during the conducts of experiments the abundance of prey declines by consumption, therefore it is necessary to take into account the depletion of prey during the duration of the experiment. To do this, Royama (1971) and Rogers (1972) separately developed equations known as "the random search equations" that allow to take into account the depletion of prey while performing the experiments. The equations assume that every prey (*N*) has the same probability of being encountered by the consumers (*P*) and that the consumer searches at random, therefore it follows the Poisson distribution.

The mathematical expression of the random search equation in Rogers (1972) is:

$$\Delta N = N \left( 1 - e^{-aP\left(T - \left(\frac{\Delta N h}{P}\right)\right)} \right)$$

For being an implicit function, since the number of prey consumed *ΔN* (=*Ne*), falls on both sides of the equation, this equation can not be solved analytically. Therefore it was necessary to use the iterative numerical method by Newton-Raphson (Juliano, 2001). Recently with the addition of the Lambert function in symbolic algebra by utilizing the software MAPLE, the analytical solution of this equation can be obtained (see Appendix I). Moreover to determine if there was depletion of prey mathematical expressions developed by Kratina et al., (2009) were used, where beginning with the standard model of Lotka-Volterra, and using the equation of the prey without production, by integrating they concluded that the functional response is related to the number of prey consumed *Ne*, by:

$$\int_{N_0}^{N_0-N_e} \frac{dN}{f(N,P,\theta)} = -\int_0^T P \, dt$$

Where *T* is the total duration of the experiment, $N_0$ is the abundance of prey at time 0 and $\theta$ are the values of the model parameters. By integrating, the explicit expressions for *Ne* are obtained (see Appendix II) which allow to estimate the parameters of the functional response in the presence of depletion of prey.

Therefore to estimate depletion during the experiments three methods were used: the numerical method of Newton-Raphson, the Rogers (1972) type equations and Kratina et al., (2009) equations (see the appendices). The Rogers type equations lead to implicit functions, the Kratina et al., equations does not. For comparison reasons the parameters were also estimated without taking into account depletion of prey.

Because some of the functional response models used in this study presented redundancy between its parameters two methods were used to check for parameter redundancy following Gimenez et al. (2004): the Hessian matrix calculation using ADMB (Fournier et al., 2012) and the formal derivative matrix using MAPLE v 13.

Hunter and Caswell (2009) refer to parameters or parameter combinations that can be estimated as estimable parameters and to parameter redundant models as rank-deficient. Determining rank deficiency is necessary for correctly interpreting model comparison measures such as Akaikie's Information Criterion (AIC) and parameter estimates. However, determining rank deficiency and the number of separably estimable parameters is a challenging problem. Until recently, there was no general and reliable way to do so (Cole et al., 2010).

Here two methods that provide estimates of redundancy were used. The first method calculates the Hessian matrix using automatic differentiation (AD) which calculates derivatives numerically, but does not use finite differences. It returns results with the same accuracy as analytical differentiation (i.e., to machine accuracy) and is more efficient than symbolic computation. This method does not distinguish between intrinsic and extrinsic redundancy, it is based on detecting the zero eigenvalues of the matrix of the second derivatives of the log–likelihood with respect to the parameters evaluated at the maximum likelihood estimates. The model rank is computed as the number of non-zero eigenvalues (i.e. the numerical rank of the Hessian is the number of linearly independent rows). The second method known as the symbolic method (Catchpole and Morgan, 1997) calculates a derivative matrix **D** and its rank. $r$ = Rank (**D**) is the number of estimable parameters $p$ in a model. The rank deficiency $d$ of the model is the difference between the number of parameters $p$ and the rank of the derivative matrix. If $d = 0$ model is full rank (not parameter redundant). If $d > 0$ model is parameter redundant. This method can distinguish between intrinsic and extrinsic redundancy, in the second case one has to take the data into account (Gimenez et al., 2004).

Finally, to compare the functional response models the Akaike Information Criterion (AIC) was used (Burnham and Anderson, 2002; Anderson, 2008; Burnham et al., 2011).

## 3. Functional response models

To discriminate between resource-dependence, consumer-dependence and ratio-dependence, and estimate whether an increasing density of consumers negatively affect per capita consumption rate, four functional response models described in Kratina al., (2009) were fitted

The models used were:

(1) Model resource-dependent, type III of Holling:

$$f(N) = \frac{N_e}{T} = \frac{aN^2}{1 + ahN^2}$$

where N is the density of resource at the beginning of the experiment, *f* is the instantaneous consumption rate (number of prey consumed per consumer per unit time), $a$[2] is the attack rate, and *h* is the handling time.

(2) Model consumer-dependent, type III of Beddington-DeAngelis, which incorporates the interference as time spent in the encounters with other consumers:

$$f(N,P) = \frac{N_e}{T} = \frac{aN^2}{1 + bw(P-1) + ahN^2}$$

where *N*, *f*, *a*, and *h* as in the previous model, *P* is the density of consumers, but the consumer-dependence is modeled as *i = b.w*, where *b* is the rate of encounters with other consumers (analogous to *a*) and *w* is the spent time with other consumers (analogous to *h*).

(3) Model ratio dependent, type III of Arditi-Ginzburg:

$$f(N,P) = \frac{N_e}{T} = \frac{a\left(\frac{N}{P}\right)^2}{1 + ah\left(\frac{N}{P}\right)^2}$$

---

[2] Also called rate of encounter with the prey or prey finding search efficiency.

(4) Model ratio and consumer dependent, type III of Arditi- Akçakaya:

$$f(N,P) = \frac{N_e}{T} = \frac{a\left(\frac{N}{P^m}\right)^2}{1 + ah\left(\frac{N}{P^m}\right)^2}$$

where *m* is the coefficient of interference (equals 0 when there is only resource-dependence and 1 when there is only ratio dependency).

## 4. Results

The logistic regression analysis determined that the curve that describes the functional relationship between the rate of consumption of *B. rotundiformis* and the density of prey, in this case *C. sorokiniana*, is a sigmoidal curve type III, because the linear terms of the third degree polynomial fitted to the data, were positive in all cases. The same result was obtained with the chemostat data of Cabrera (2011) (Table 1). This means that the effectiveness at which *Chlorella* is consumed has a maximum at intermediate values of algal density. The per capita consumption rate of consumers decreased by increasing the density of co-specifics and this effect was particularly evident at high densities of prey (Figs. 1 and 2). The Table 2 shows the results of nonlinear fitting of the data of the consumption rate without taking into account the depletion of the resource vs. the resource density, with different sigmoidal models of functional response (Holling III, Beddington-De Angelis III, Arditi-Ginzburg III, Arditi-Akçakaya III). When analyzing the behavior of the parameter values relative to the increase in the density of consumers without depletion of the resource we observed that only the Holling III model was able to predict that with an increase in consumer density (interference), the rate of encounters with the resource decreases, and therefore its consumption. Whereas the other models (Beddington-De Angelis III, Arditi-Ginzburg III, Arditi-Akçakaya III) predict the opposite.

**Table 1**. Results of logistic regressions of proportion of *Chlorella sorokiniana* consumed by *Brachionus rotundiformis* and number of offered *Chlorella*. Positive linear terms indicate a Type III functional response. In the chemostat the same is obtained.

| Functional response experiments | | |
|---|---|---|
| Consumer density (ind/ml) | Linear term | Standard error |
| 1 | 4,95E-8 | 5,19E-9 |
| 5 | 2,6E-7 | 4,46E-9 |
| 25 | 1,25E-6 | 4,38E-9 |
| 125 | 1,78E-7 | 5,0E-9 |
| All data | 3,01E-6 | 9,9E-10 |
| Functional response in the chemostat (Cabrera, 2011) | | |
| All data | 6,38E-06 | 5,41E-09 |

On the other hand, when the depletion of the resource is taken into account (see Table 3 and Figure 3) it is observed that the Holling III model does not converged at high densities of the consumers, whereas the Arditi-Ginzburg III model, fluctuates in its encounter rate as the density of the consumers increases, only the Beddington-De Angelis III and the Arditi-Akçakaya III models decrease the rate of encounters with resource as the density of consumers increase (Figure 4). In all models the handling time increases as the density of consumers increases and the depletion is less.

To compare the models, the Akaike Information Criterion (AIC) was used. Table 2 shows that without taking into account depletion of the resource, all the models have the same sum of squares (SS) and the same values for AICc for all densities of consumers. Therefore, all have the same probability of being correct, (ie since the AICc values are the same, then the differences ΔAIC are zero, and each model is equally likely to be correct). Moreover, there is a 50% chance for each model without depletion to be correct and hence 50% chance that is incorrect. However, taking into account depletion of resource (Table 3), it is observed first that the sigmoidal model of Holling does not converged for values of P > 5, and when P = 1, the models that incorporate the effect of depletion, had the smallest information loss because their ΔAIC's were negative. As covariance between the encounter rate *a* and the interference parameter *i*, in the Beddington-De Angelis model was one, and covariance between the encounter rate *a* and the interference parameter *m* in the Arditi-Akçakaya model was also one, the nonlinear regression method excludes both *i* as *m*, because of co-dependency with *a* . This means that both models could be redundant in their parameters, ie fewer parameters or parameter combinations can be estimated than the original number of parameters in the models. To verify that parameter redundancy was not related to sample size or the number of replicates, it was proceeded to generate data using bootstrapping. Four sets of 150 points each were generated from the original data. Each of these data sets were analyzed in a similar way to the original data, obtaining similar results (see Table 4).

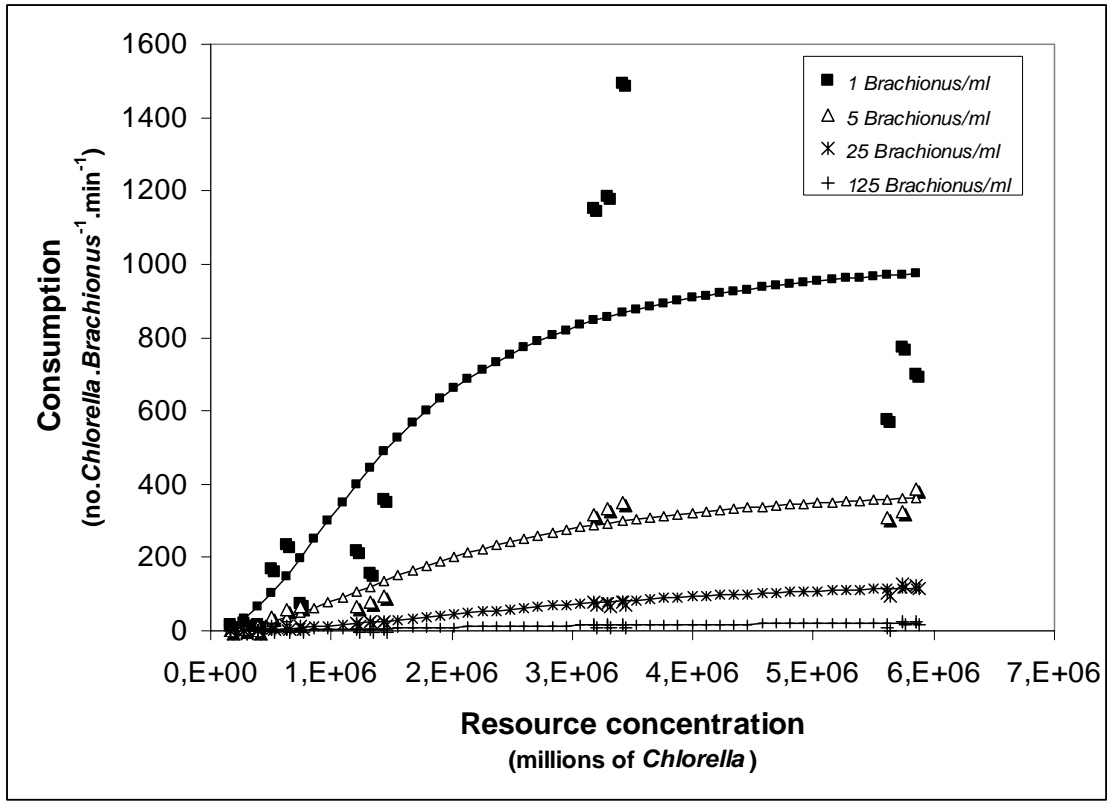

**Figure 1.** Consumption rates of *B. rotundiformis* at four different consumer concentrations. Lines are fitted curves for Type III functional response for each consumer concentration.

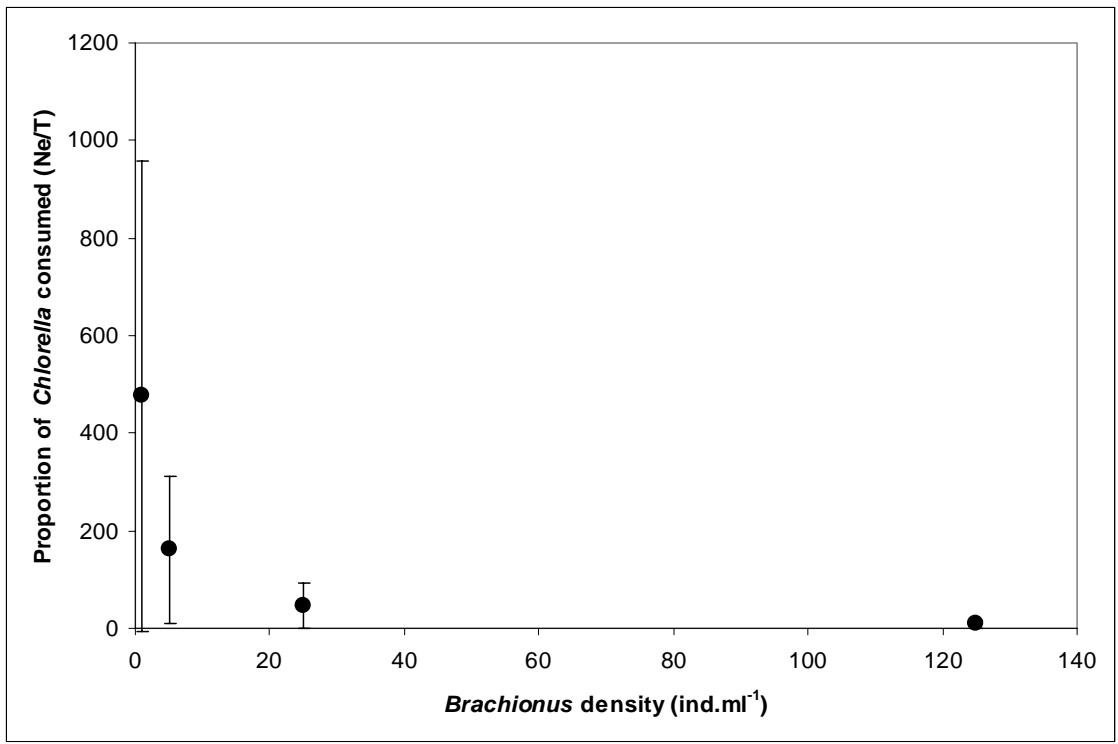

**Figure 2.** Effect of the density of *B. rotundiformis* on proportion of *Chlorella* consumed. The mean ± 1 SD of each density is shown. Proportion of prey consumed declined with the increase in density of co-specifics.

To demonstrate the redundancy of parameters in the Beddington - De Angelis and Arditi-Akçakaya models, in a first approach the Hessian matrix was calculated using automatic differentiation (AD) through the AD Model Builder (ADMB) software (see Table 5). As was expected the Hessian matrix was not positive definite i.e. inability to find the minimum of the objective function thus signaling a perfect linear dependency between *a* and *i* and *a* and *m*.

The Hessian matrix only was positive definite when $i = m = 0$ and when constraints on "*a*" were made, making it constant, allowing only the simultaneous estimation of *h* and *i* and *h* and *m*.

The second approach to demonstrate the redundancy of parameters in the Beddington-De Angelis and Arditi-Akçakaya models was by using the symbolic method. In the Beddington - De Angelis model, ***d*** =1 > 0, i.e. the model is intrinsically parameter redundant resulting from the structure of the model and irrespective of the data.

**Table 2. The parameter estimates (±1 SE) for the fits of four functional response models without depletion to de data for 1 to 125 consumers.**

| Model Without depletion | Encounter rate a (with prey.min$^{-1}$) | Handling time h (min$^{-1}$) | Interference parameter [a] | K | AICc[b] | ΔAICc[c] |
|---|---|---|---|---|---|---|
| **1 consumer** | | | | | | |
| Beddington-DeAngelis III | 4,4e-10± 2,5e-10 | 0,0009600± 0,0001558 | $i$[d] = 0 | 2 | 171.4306952 | 0 |
| | | | | 3 | 174.6125134 | 3.18 |
| Arditi-Akçakaya III | 4,4e-10± 2,5e-10 | 0,0009600± 0,0001558 | $m$ = 0 | 2 | 171.4306952 | 0 |
| | | | | 3 | 174.6125134 | 3.18 |
| Arditi-Ginzburg III | 4,4e-10± 2,5e-10 | 0,0009600± 0,0001558 | | 2 | 171.4306952 | 0 |
| Holling III | 4,4e-10± 2,5e-10 | 0,0009600± 0,0001558 | | 2 | 171.4306952 | 0 |
| **5 consumers** | | | | | | |
| Beddington-DeAngelis III | 8,5e-10± 1,7e-10 | 0,002470± 0,0001567 | $i$ =1,911±0,3073 | 2 | 106,92 | 0 |
| | | | | 3 | 110,0972571 | 3.18 |
| Arditi-Akçakaya III | 8,5e-10± 1,7e-10 | 0,002470± 0,0001567 | $m$ = 0,6709 | 2 | 106,92 | 0 |
| | | | | 3 | 110,0972571 | 3.18 |
| Arditi-Ginzburg III | 2,5e-9± 4,9e-10 | 0,002470± 0,0001567 | | 2 | 106,92 | 0 |
| Holling III | 9,9e-11± 1,9e-11 | 0,002470± 0,0001567 | | 2 | 106,92 | 0 |
| **25 consumers** | | | | | | |
| Beddington-DeAngelis III | 1,7e-9± 2,3e-10 | 0,006715± 0,0004363 | $i$ = 4,539±0,3500 | 2 | 59,94 | 0 |
| | | | | 3 | 63,12498296 | 3.18 |
| Arditi-Akçakaya III | 1,7e-9± 2,3e-10 | 0,006715± 0,0004363 | $m$ = 0,7228 | 2 | 59,94 | 0 |
| | | | | 3 | 63,12498296 | 3.18 |
| Arditi-Ginzburg III | 9,7e-9± 1,3e-9 | 0,006715± 0,0004363 | | 2 | 59,94 | 0 |
| Holling III | 1,6e-11± 2,1e-12 | 0,006715± 0,0004363 | | - | - | - |
| **125 consumers** | | | | | | |
| Beddington-DeAngelis III | 1,1e-8± 4,9e-9 | 0,04139± 0,006939 | $i$ =22,16±6,366 | 2 | 43,94 | 0 |
| | | | | 3 | 47,11809962 | 3.18 |
| Arditi-Akçakaya III | 1,1e-8± 4,9e-9 | 0,04139± 0,006939 | $m$ = 0,8246 | 2 | 43,94 | 0 |
| | | | | 3 | 47,11809962 | 3.18 |
| Arditi-Ginzburg III | 6,2e-8± 2,8e-8 | 0,04139± 0,006939 | | 2 | 43,94 | 0 |
| Holling III | 4,0e-12± 1,8e-12 | 0,04139± 0,006939 | | - | - | - |

a cov ($a,i$) and cov ($a,m$) = 1, K Number of parameters, b AICc Adjusted Akaike's information criterion, c ΔAICc The difference from the best model, d (encounters with consumer.min$^{-1}$).

**Table 3.** The parameter estimates (±1 SE) for the fits of four functional response models with depletion to de data for 1 to 125 consumers.

| Model With depletion | Encounter rate a (with prey.min$^{-1}$) | Handling time h (min$^{-1}$) | Interference parameter [a] | K | AICc[b] | ΔAICc[c] |
|---|---|---|---|---|---|---|
| **1 consumer** | | | | | | |
| Beddington-DeAngelis III | 1,478e-011± 8,474e-012 | 0,02880± 0,004674 | $i^d$ = 0 | 2 | 171,4305585 | -0,0001366414 |
| | | | | 3 | 174,6123767 | 3,1818182 |
| Arditi-Akçakaya III | 1,478e-011± 8,474e-012 | 0,02880± 0,004674 | $m$ = 0 | 2 | 171,4305585 | -0,0001366414 |
| | | | | 3 | 174,6123767 | 3,1818182 |
| Arditi-Ginzburg III | 1,478e-011± 8,474e-012 | 0,02880± 0,004674 | | 2 | 171,4305585 | -0,0001366414 |
| Holling III | 1,478e-011± 8,474e-012 | 0,02880± 0,004674 | | 2 | 171,4305585 | -0,0001366414 |
| **5 consumers** | | | | | | |
| Beddington-DeAngelis III | 6,688e-012± 1,317e-012 | 0,3705± 0,02351 | $i$ = 2,288±0,4996 | 2 | 106,92 | 0 |
| | | | | 3 | 110,0972571 | 3,1818182 |
| Arditi-Akçakaya III | 6,687e-012± 1,316e-012 | 0,3705± 0,02351 | $m$ = 0,6709 | 2 | 106,92 | 0 |
| | | | | 3 | 110,0972571 | 3,1818182 |
| Arditi-Ginzburg III | 1,647e-011± 3,242e-012 | 0,3705± 0,02351 | | 2 | 106,92 | 0 |
| Holling III | 6,587e-013± 1,297e-013 | 0,3705± 0,02351 | | 2 | 106,92 | 0 |
| **25 consumers** | | | | | | |
| Beddington-DeAngelis III | 2,278e-012± 3,047e-013 | 5,036± 0,3272 | $i$ = 4,537±0,6124 | 2 | 59,94 | 0 |
| | | | | 3 | 63,12498296 | 3,1818182 |
| Arditi-Akçakaya III | 2,278e-012± 3,047e-013 | 5,036± 0,3272 | $m$ = 0,7228 | 2 | 59,94 | 0 |
| | | | | 3 | 63,12498296 | 3,1818182 |
| Arditi-Ginzburg III | 1,296e-011± 1,733e-012 | 5,036± 0,3272 | | 2 | 59,94 | 0 |
| Holling III | Not converged | Not converged | | - | - | - |
| **125 consumers** | | | | | | |
| Beddington-DeAngelis III | 2,916e-012± 1,293e-012 | 155,2± 26,02 | $i$ = 22,15±9,824 | 2 | 43,94 | 0 |
| | | | | 3 | 47,11809962 | 3,1818182 |
| Arditi-Akçakaya III | 2,916e-012± 1,293e-012 | 155,2± 26,02 | $m$ = 0,8246 | 2 | 43,94 | 0 |
| | | | | 3 | 47,11809962 | 3,1818182 |
| Arditi-Ginzburg III | 1,659e-011± 7,354e-012 | 155,2± 26,02 | | 2 | 43,94 | 0 |
| Holling III | Not converged | Not converged | | - | - | - |

a cov (*a,i*) and cov (*a,m*) = 1, K Number of parameters, b AICc Adjusted Akaike's information criterion, c ΔAICc The difference from the best model, d (encounters with consumer.min$^{-1}$).

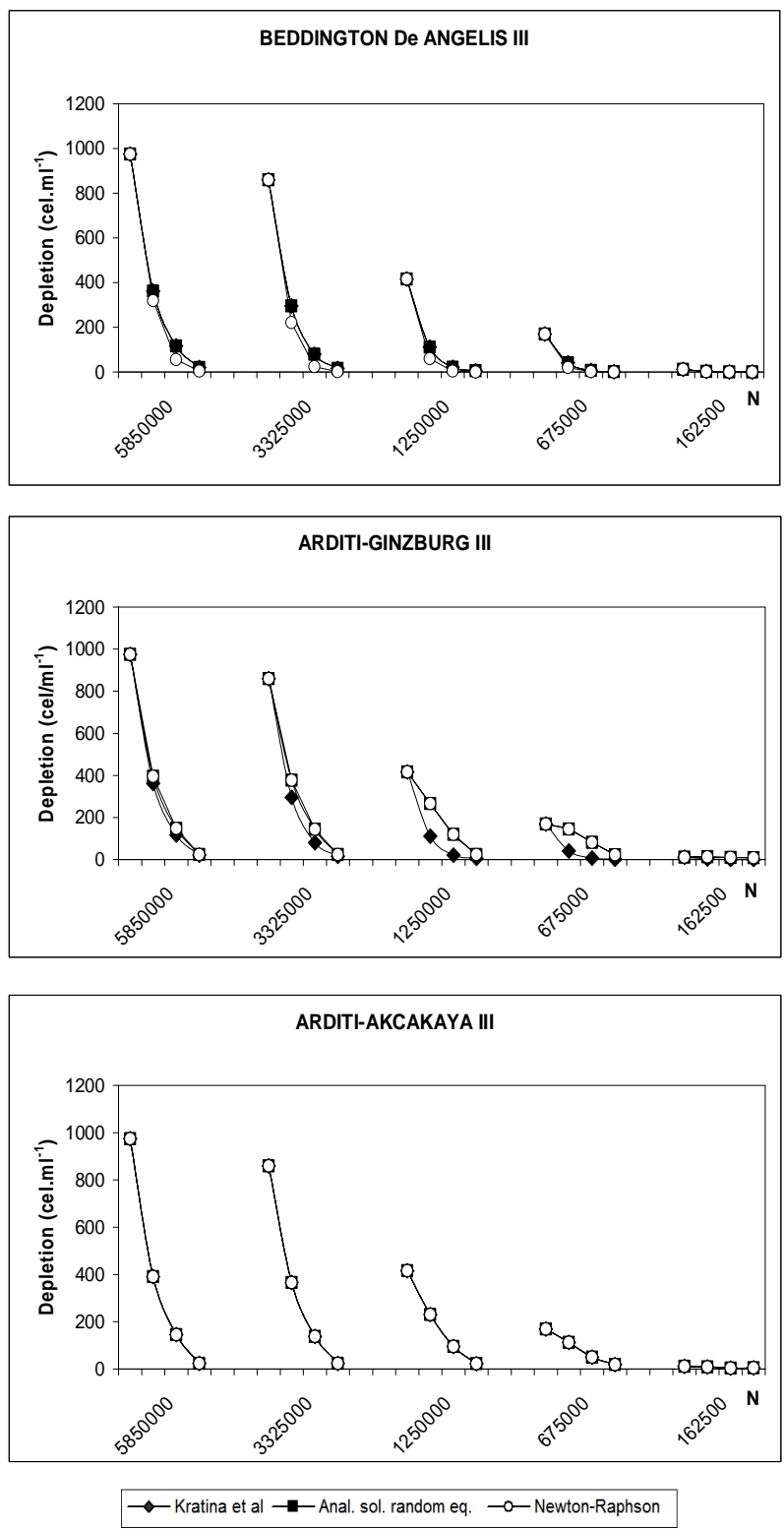

**Figure 3.** Depletion values during the experiments. The results of the three methods that were used to estimate it (Newton-Raphson, Random equations, Kratina et al., equations) were similar.

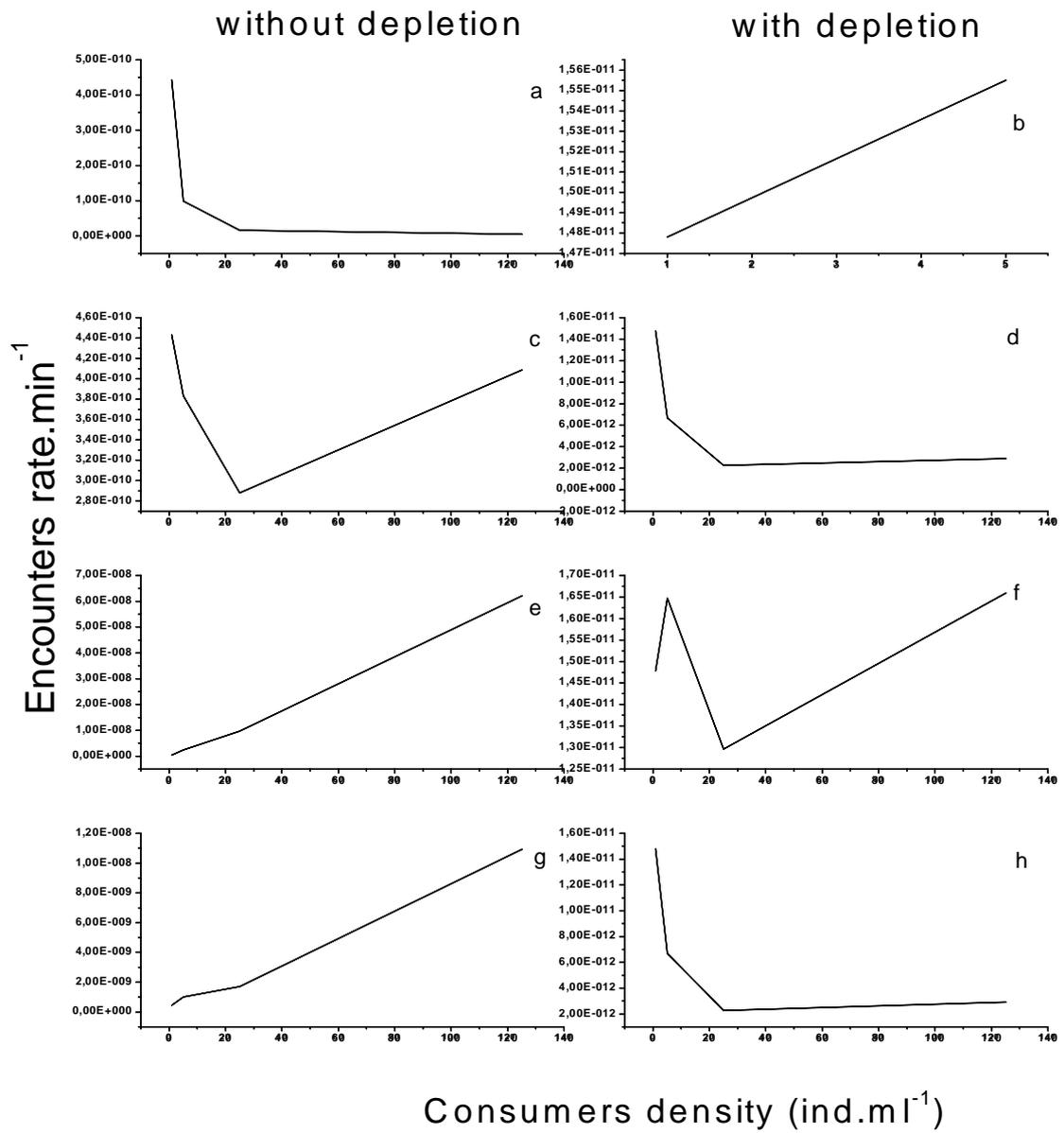

**Figure 4.** Behaviour of the encounter rate parameter in relation to consumer density without depletion and with depletion. a,b: Holling model; c,d: Beddington-DeAngelis model; e,f: Arditi-Ginzburg model; g,h: Arditi-Akçakaya model.

**Table 4. Four sets of 150 points were generated from the original data using bootstrapping to verify that parameter redundancy was not related to sample size or the number of replicates. The parameter estimates are similar to the original data (see text for more details).**

| Beddinton-DeAngelis-III *with depletion* | | | |
|---|---|---|---|
| **Best-fit values** | 5 ind.ml$^{-1}$ | 25 ind.ml$^{-1}$ | 125 ind.ml$^{-1}$ |
| *a* | ~ 6.688e-012 ± ~ 1.036e-010 | ~ 2.387e-012 ± ~ 2.658e-012 | ~ 2.898e-012 ± ~ 2.185e-013 |
| *h* | 0,3705 ± 1,260e-007 | 5,036±8,976e-007 | 155,2 ± 6,420e-006 |
| *i* | ~ 2.288 ± ~ 39.32 | ~ 4.757±~ 5.343 | ~ 22.01 ± ~ 1.660 |
| **Covariance Matrix** | | | |
| *i* and *a* | 1,000 | 1,000 | 1,000 |
| *i* and *h* | 0,1072 | -0,09932 | -0,04284 |
| *a* and *h* | 0,1072 | -0,09932 | -0,04284 |
| **Dependency** | | | |
| *a* | 1,000 | 1,000 | 1,000 |
| *h* | 0,5881 | 0,7169 | 0,6476 |
| *i* | 1,000 | 1,000 | 1,000 |
| **Number of points** | 150 | 150 | 150 |

| Arditi-Akçakaya -III *with depletion* | | | |
|---|---|---|---|
| **Best-fit values** | 5 ind.ml$^{-1}$ | 25 ind.ml$^{-1}$ | 125 ind.ml$^{-1}$ |
| *a* | ~ 5.648e-013 ± ~ 2.444e-012 | ~ 1.333e-012 ± ~ 1.223e-012 | ~ 2.908e-012 ± ~ 3.480e-013 |
| *h* | 0,3705 ± 1,254e-007 | 5,036 ± 8,945e-007 | 155,2 ± 6,414e-006 |
| *m* | ~ -0.04782 ± ~ 1.344 | ~ 0.6467 ± ~ 0.1425 | ~ 0.8197 ± ~ 0.01239 |
| **Covariance Matrix** | | | |
| *m* and *a* | 1,000 | 1,000 | 1,000 |
| *m* and *h* | 0,04869 | 0,05507 | -1,861e-021 |
| *a* and *h* | 0,04869 | 0,05507 | 6,320e-007 |
| **Dependency** | | | |
| *a* | 1,000 | 1,000 | 1,000 |
| *h* | 0,5843 | 0,715 | 0,6469 |
| *m* | 1,000 | 1,000 | 1,000 |
| **Number of points** | 150 | 150 | 150 |

The resulting expression for the derivative involves the parameters with respect to which one is differentiating. This led to an excessive growth of the length of the expression making a reparameterization impossible. Figure 5 shows the rank of the derivative matrix of the Beddington-De Angelis model. In the Arditi- Akçakaya model, it was not possible to calculate the derivative matrix **D** because the length of the output exceeds the limit 1000000, thus exceeding the capabilities of the symbolic math software.

**Table 5. Results of the parameter estimation from the Beddington-DeAngelis III and Arditi-Akçakaya III functional response models using Automatic Differentiation (AD).**

| Beddington-DeAngelis III | | | | | | | |
|---|---|---|---|---|---|---|---|
| **(a, h, i)** | | | | | | | |
| Consumer's number | *a* | *h* | *i* | cov | Objetive function | Maximum gradient | Log Determinant of Hessian |
| 1 | 1.60e-011 ± 3.86e-012 | 3.50e-002 ± 7.98e-003 | 0 | 0.37 | 9.64 | 12.71 | 62.37 |
| 5* | 5.21e-012 | 0.38 | 1.82 | ? | -0.10 | 85172.8 | - |
| 25* | 2.63e-012 | 5.999 | 2.5 | ? | 3.17 | 201.53 | - |
| 125* | 4.93e-012 | 200.0 | 23.0 | ? | 3.88 | 138.72 | - |
| **(a, i)** | | | | | | | |
| 5* | 6.08e-012 | 0.5 | 2.0 | 1.0 | 2.07 | 20.75 | - |
| 25* | 2.48e-012 | 5.0 | 2.5 | 1.0 | 4.72 | 534.8 | - |
| 125* | 4.6e-012 | 160.0 | 23.0 | 1.0 | 4.90 | 96.28 | - |
| **(h, i)** | | | | | | | |
| 5 | 6.0e-012 | 0.38±0.045 | 2.13 ± 0.24 | -0.33 | -0.10 | 7.65e-009 | 8.96 |
| 25 | 2.28e-012 | 6.0±0.4315 | 2.16 ± 0.27 | -0.0004 | 3.17 | 0.0013 | 18.11 |
| 125 | 2.92e-012 | 200.0 ± 1.25e-002 | 13.6 ± 1.8 | 0.0021 | 3.88 | 4.53e-006 | 7.58 |
| **Arditi-Akçakaya III** | | | | | | | |
| **(a, h, m)** | | | | | | | |
| Consumer's number | *a* | *h* | *m* | cov | Objetive function | Maximum gradient | Log Determinant of Hessian |
| 1 | 1.60e-011 ± 3.86e-012 | 3.50e-002 ± 7.98e-003 | 0 | 0.37 | 9.64 | 10.89 | 62.37 |
| 5* | 3.15e-012 | 0.38 | 0.5 | ? | -0.10 | 464.80 | - |
| 25* | 1.14e-012 | 7.24 | 0.5 | ? | 2.50092 | 282.93 | - |
| 125* | 2.16e-013 | 200.0 | 0.5 | ? | 3.88 | 10933.0 | - |
| **(a, m)** | | | | | | | |
| 5* | 3.20e-012 | 0.40 | 0.5 | 1.0 | 0.02 | 533.94 | - |
| 25* | 1.02e-012 | 5.0 | 0.5 | 1.0 | 4.72 | 1473.9 | - |
| 125* | 2.02e-013 | 160.0 | 0.5 | 1.0 | 4.90 | 3887.42 | - |
| **(h, m)** | | | | | | | |
| 5 | 6.69e-012 | 0.38±0.045 | 0.73 ± 0.03 | -0.33 | -0.10 | 2.32e-005 | 13.03 |
| 25 | 2.28e-012 | 7.24± 1.12 | 0.61 ± 0.02 | -0.38 | 2.5 | 1.21e-005 | 7.73 |
| 125 | 2.92e-012 | 200.0±3.34e-002 | 0.77 ± 0.01 | -0.003 | 3.88 | 3.31e-004 | 15.38 |

*****Hessian does not appear to be positive definite.**

*a* **(Encounter rate)**
*h* **(Handling time)**
*i, m* **(Interference parameters)**

$$r := \text{Rank}\left(\text{Dmat}\left(\begin{bmatrix} 0 \\ \dfrac{2\,a^2 N0^3}{(1+i(P-1)+ahN0)^2} - \dfrac{a^3 N0^4 h}{(1+i(P-1)+ahN0)^3} \\ \dfrac{6\,a^3 N0^4}{(1+i(P-1)+ahN0)^3} - \dfrac{8\,a^4 N0^5 h}{(1+i(P-1)+ahN0)^4} + \dfrac{3\,a^5 N0^6 h^2}{(1+i(P-1)+ahN0)^5} \\ \dfrac{24\,a^4 N0^5}{(1+i(P-1)+ahN0)^4} - \dfrac{58\,a^5 N0^6 h}{(1+i(P-1)+ahN0)^5} + \dfrac{50\,a^6 N0^7 h^2}{(1+i(P-1)+ahN0)^6} - \dfrac{15\,a^7 N0^8 h^3}{(1+i(P-1)+ahN0)^7} \\ \dfrac{120\,a^5 N0^6}{(1+i(P-1)+ahN0)^5} - \dfrac{444\,a^6 N0^7 h}{(1+i(P-1)+ahN0)^6} + \dfrac{640\,a^7 N0^8 h^2}{(1+i(P-1)+ahN0)^7} - \dfrac{420\,a^8 N0^9 h^3}{(1+i(P-1)+ahN0)^8} + \dfrac{105\,a^9 N0^{10} h^4}{(1+i(P-1)+ahN0)^9} \end{bmatrix}, \begin{bmatrix} N0 \\ a \\ i \\ h \\ P \end{bmatrix}\right)\right)$$

**Figura 5.** Rank of the derivative matrix (Dmat) of the Beddington-DeAngelis model using MAPLE v 13 symbolic algebra software.

Because of the rank deficiency it is therefore not possible to correctly interpret model comparison measures such as Akaike's Information Criterion (AIC) and parameter estimates. However, discarding the results that do not take into account the depletion of resource and considering the redundancy of parameters, models incorporating intraspecific interference, even at low densities of consumers, are the ones that best describe the Type III functional response of *B. rotundiformis*.

## 5. Discussion

The results of the present study suggest that the functional response of the filter feeder *B. rotundiformis* is of Type III. This result agrees with those found for other filter feeders as *B. calyciflorus* and *Daphnia* (Fussman et al., 2005; Sarnelle and Wilson, 2008) that were traditionally located within the functional response Type I and II (Jeschke et al., 2004). After Holling (1959b) the different types of functional response were associated with the mode and complexity of consumer foraging. Thus Type I was associated with invertebrate filtering feeders, Type II with invertebrate not filtering feeders and Type III with higher organisms (vertebrates) (Murdoch and Oaten, 1975; May, 1981). However, Hassell et al., (1977) pointed out that the Type III functional response could not be confined only to vertebrate consumers and they emphasized the need for experiments at low density of resources. There are two reasons that explain why Type I and fundamentally Type II have prevailed as the types of functional response most commonly encountered for invertebrates. First, due to data not covering the entire range of possible concentrations of food, including low densities of food and secondly, because the statistical methodology used did not properly allow the detection of the data pattern at low food densities. Previous work by Trexler et al., (1988) and Trexler and Travis (1993) has established that many functional responses that thought to be Type II have proved to be Type III, as Hassell et al., (1977) had pointed out long ago.

The importance of distinguishing between Type II and Type III functional responses has to do with their very different contributions to the stability of the consumer- resource systems. The sigmoidal Type III functional response is density dependent up to a threshold of resource density and can contribute to the stability if the average density of resource is below this threshold (Hassell et al., 1977). The effect of a sigmoidal functional response on the stability of phytoplankton-zooplankton interaction has been discussed by Steele (1974) and in a broader context by Murdoch and Oaten (1975).

Another fundamental difference between Type II and Type III functional response is that in the former the encounter rate with resource or attack rate "$a$" remains constant while in the latter it varies with the density of resource. In this study the parameter estimates of the encounter rate "$a$", the handling time "$h$" and the interference competition parameter "$i$" in the case of the Beddington-DeAngelis model or "$m$" in the case of Arditi-Akçakaya model were all estimated without depletion (instantaneous consumption rate) and with depletion (integrated consumption rate) of resource during the experiments for comparison reasons. The results are not the same (Tables 2, 3 and Figure 4). When depletion is not considered, the encounter rate estimates increases in general with consumer density, however, when depletion is taken into account only the Beddington-DeAngelis III and the Arditi-Akçakaya III models decrease the rate of encounters with resource as the density of consumers increase. This contradictory result reflects the importance of taking into account depletion in experiments where there is no replacement of resource. Obviously, the correct estimates of the parameters were those obtained by taking into account depletion. As consumer density increases consumers encounter rate "$a$" decrease and handling time "$h$" increases as a result of intraspecific competition by interference ("$i$" or "$m$"). However, other studies found no difference between the instantaneous consumption rate and integrated consumption rate (Fussman et al., 2005; Jost et al., 2005) probably because the experiments were very short.

Another result, which I think is the most important and worrying of this study, is the identifiability problem or redundancy of model parameters incorporating intraspecific competition by interference. As Paine (1988) argued, for the theory of food webs become predictive in a dynamical sense, it should incorporate competitive interactions in consumer-resource interactions. The fact that fewer parameters or parameter combinations can be estimated than the original number of parameters in consumer-resource models implies that certain constraints on parameters in the model need to be impose (Fujiwara and Caswell, 2002; Kendall and Nichols, 2002). Parameter constraints can represent a priori information about species biology or biological hypotheses to be tested by model selection. To the extent that the constraints are biologically reasonable, this procedure can be a valuable approach in constructing and testing biological hypotheses. However whenever a model is not uniquely identifiable, there are several parameter values that correspond to exactly the same input-output behavior; and the very meaning of an attempt to estimate them is questionable (Walter, 1987). The problem is very simple to explain but no one can claim until now to have solved it definitively. When model parameters are not identifiable, one has little confidence that estimated values are close to the true values.

Here, the model parameters that incorporate intra-specific competition by interference in the consumer-resource interaction are not identifiable, suggesting that this problem may be more widespread than is generally appreciated in the literature of food webs.

However, parameter non-identifiability in biochemical systems was investigated since the 1970s (Reich, 1974; Reich et al., 1974a,b; Reich and Zinke, 1974; Reich, 1981) and more recently by Ashyraliyev et al (2009). The same has happened in microbiological systems. As Jost (1998) posits, consumption of a resource by an organism is a key process in both microbiology and population ecology. Gutenkunst et al., (2007) found that for all 17 systems biology models that they considered, the obtained parameters are 'sloppy', meaning not well-defined. They argued that sloppiness emerges from a redundancy between the effects of different parameter combinations. In ecology, capture-recapture models show time-dependence of both survival and capture probabilities and it is not possible to estimate the individual components (Newman et al., 2014).

In this study, the obtained results show that intra-specific interference is important to correctly explain the process of consumption but it can not be quantified individually with reasonable accuracy due to codependence with encounter rate. Because proper estimation of model parameters is required for ensuring accurate model predictions this result is a matter of concern, considering that the interference interaction strength of functional responses is crucially important for the stability of consumer–resource systems and the persistence, sustainability and biodiversity of ecosystems. The degree of intra-specific interference determines the rate of encounters and handling time. To the extent that the interference increases, the rate of encounter with the prey (consumption) decreases, which increases the time of handling the prey. The encounter rate and handling time is determined by the interference. The rate of encounter with the prey and handling time may only take values that are inherently determined by the degree of interference. Therefore, as consumers increased in density, intra-specific interference was increased, and consumption was decreased.

Including knowledge on competitive interactions in current model predictions will be a necessity in the next years for ecology as ecological models are getting more complex and more real. Identifiability of biological parameters in nonlinear ecological models will be an issue to consider.

## 6. Conclusions

The functional response of *B. rotundiformis* is a Type III response.

The rate of consumption of *B. rotundiformis* is consumer dependent even at relatively low consumer densities.

The degree of interference product of the density of consumers determines the rate of encounters and handling time. To the extent that the interference increases, the rate of encounter with the prey (consumption) decreases, which increases the time of handling the prey. The encounter rate and handling time is determined by the interference. The rate of encounter with the prey and handling time may only take values that are inherently determined by the degree of interference. Therefore, as consumers increased in density, intra-specific interference was increased, and consumption was decreased.

The Beddington-DeAngelis and Arditi-Akçakaya Type III functional response models are non-identifiable or parameter redundant because of a perfect linear dependency between $a$ (encounter rate) and $i$ (interference parameter) in the former case and $a$ (encounter rate) and $m$ (interference parameter) in the latter.

# Appendix I

a. Analitical solution of the Rogers random implicit equation for more than one predator (eq. 22 in Rogers, 1972) using the Lambert function of MAPLE v13:

$$\Delta N = N\left(1 - e^{-aP\left(T - \left(\frac{\Delta Nh}{P}\right)\right)}\right) \xrightarrow{isolate\ for\ \Delta N}$$

$$\Delta N = \frac{ahN - LambertW\left(ahNe^{-PTa + ahN}\right)}{ah}$$

Here $W$ is the Lambert $W$ function, defined as the real valued solution of the following equation (Corless *et al.*, 1996):

$$W(x)e^{W(x)} = x$$

b. Random equations for type III functional responses developed in this work by analogy with the equation 22 in Rogers (1972) and their analytical solutions using the Lambert function of MAPLE v13:

1-Holling III

$$\Delta N = N\left(1 - e^{-aNP\left(T - \left(\frac{\Delta Nh}{P}\right)\right)}\right) \xrightarrow{isolate\ for\ \Delta N}$$

$$\Delta N = \frac{ahN^2 - LambertW\left(ahN^2 e^{-aNPT + ahN^2}\right)}{ahN}$$

2-Arditi-Ginzburg III

$$\Delta N = N\left(1 - e^{-a\left(\frac{N}{P}\right)P\left(T - \left(\frac{\Delta Nh}{P}\right)\right)}\right)$$

$$\Delta N = N\left(1 - e^{-aN\left(T - \left(\frac{\Delta Nh}{P}\right)\right)}\right) \xrightarrow{isolate \cdot for \cdot \Delta N}$$

$$\Delta N = \frac{ahN^2 - P.LambertW\left(\frac{ahN^2 e^{\frac{Na(-PT + hN)}{P}}}{P}\right)}{ahN}$$

3-Arditi-Akcacaya III

$$\Delta N = N\left(1 - e^{-a\left(\frac{N}{P^m}\right)P\left(T - \left(\frac{\Delta Nh}{P}\right)\right)}\right)$$

$$\Delta N = N\left(1 - e^{-\frac{aNP\left(T - \frac{\Delta Nh}{P}\right)}{P^m}}\right) \xrightarrow{isolate \cdot for \cdot \Delta N}$$

$$\Delta N = \frac{ahN^2 - P^m LambertW\left(ahN^2 e^{Na(-PT + hN)P^{-m}} P^{-m}\right)}{ahN}$$

4-Bedington-de Angelis III (to derive this expression the work of Beddington (1975) was used).

$$\Delta N = N\left(1 - e^{\left(\frac{aN(\Delta Nh - PT)}{1 + bw(P - 1)}\right)}\right)$$

$$\Delta N = \frac{1}{aNh} \left( \begin{array}{l} -\left( -LambertW\left( -\frac{aN^2 he^{\frac{aN(Nh-PT)}{1+bwP-bw}}}{bw-1-bwP} \right) + \frac{aN(Nh-PT)}{1+bwP-bw} \right)bw + aNPT - \\ LambertW\left( -\frac{aN^2 he^{\frac{aN(Nh-PT)}{1+bwP-bw}}}{bw-1-bwP} \right) + \frac{aN(Nh-PT)}{1+bwP-bw} + \\ \left( -LambertW\left( -\frac{aN^2 he^{\frac{aN(Nh-PT)}{1+bwP-bw}}}{bw-1-bwP} \right) + \frac{aN(Nh-PT)}{1+bwP-bw} \right)bwP \end{array} \right)$$

# Appendix II

Equations for type III functional responses using methodology from Arditi-Saiah (1992) and Jost et al, (2005) (see Kratina et al,2009)

Holling, type III

$$\left( \frac{\left(ah(N-\Delta N)-\left(\frac{1}{N-\Delta N}\right)\right)}{a} - \frac{\left(ahN-\left(\frac{1}{N}\right)\right)}{a} + PT = 0 \right)$$

simplify

$$\frac{-ahN^2\Delta N + Nah\Delta N^2 - \Delta N + PTN^2 a - PTNa\Delta N}{Na(N-\Delta N)} = 0$$

isolate for $\Delta N$

$$\Delta N = \frac{1}{2}\frac{1}{ahN}\left(ahN^2 + 1 + PTNa - \left(a^2 h^2 N^4 + 2ahN^2 - 2a^2 hN^3 PT + 1 + 2PTNa + P^2 T^2 N^2 a^2\right)^{1/2}\right)$$

For factorizing the polynomial inside the square root the term $4TPha^2 N^3$ is added.

$$Po = a^2 h^2 N^4 + 2ahN^2 - 2a^2 hN^3 PT + 1 + 2PTNa + P^2 T^2 N^2 a^2 + 4TPha^2 N^3$$

$$Po = a^2 h^2 N^4 + 2ahN^2 + 2a^2 hN^3 PT + 1 + 2PTNa + P^2 T^2 N^2 a^2$$

$factor(Po);$

$$\left(ahN^2 + 1 + PTNa\right)^2$$

and

$$\Delta N = \frac{ahN^2 + 1 + PTNa - \sqrt{\left(ahN^2 + 1 + PTNa\right)^2 - 4TPha^2 N^3}}{2ahN}$$

Beddington-de Angelis, type III

$$\left( \frac{\left(ah(N-\Delta N)-\left(\frac{1+bwP-bw}{N-\Delta N}\right)\right)}{a} - \frac{\left(ahN-\left(\frac{1+bwP-bw}{N}\right)\right)}{a} + PT = 0 \right)$$

simplify

$$\frac{-ahN^2\Delta N + Nah\Delta N^2 - \Delta N + bwP\Delta N + bw\Delta N + PTN^2a - PTNa\Delta N}{Na(N-\Delta N)} = 0$$

Isolate for ΔN

$$\Delta N = \frac{1}{2}\frac{1}{ahN}\left( ahN^2 + 1 + bwP - bw + PTNa - \begin{pmatrix} 1 - 2b^2w^2P + 2ahN^2bwP + 2bwP^2TNa - 2bwPTNa \\ -2bw - 2a^2hN^3PT - 2ahN^2bw + 2PTNa + 2ahN^2 \\ + P^2T^2N^2a^2 + a^2h^2N^4 + 2bwP + b^2w^2P^2 + b^2w^2 \end{pmatrix}^{1/2} \right)$$

For factorizing the polynomial inside the square root the term $4TPha^2N^3$ is added.

Po =

$1 - 2b^2w^2P + 2ahN^2bwP + 2bwP^2TNa - 2bwPTNa$
$- 2bw - 2a^2hN^3PT - 2ahN^2bw + 2PTNa + 2ahN^2$
$+ P^2T^2N^2a^2 + a^2h^2N^4 + 2bwP + b^2w^2P^2 + b^2w^2 + 4TPha^2N^3$

Po =

$1 - 2b^2w^2P + 2ahN^2bwP + 2bwP^2TNa - 2bwPTNa$
$- 2bw + 2a^2hN^3PT - 2ahN^2bw + 2PTNa + 2ahN^2$
$+ P^2T^2N^2a^2 + a^2h^2N^4 + 2bwP + b^2w^2P^2 + b^2w^2$

$factor(Po);$

$(ahN^2 + 1 + bwP - bw + PTNa)^2$

and

$$\Delta N = \frac{ahN^2 + 1 + bw(P-1) + PTNa - \sqrt{(ahN^2 + 1 + bw(P-1) + PTNa)^2 - 4TPha^2N^3}}{2ahN}$$

Arditi-Ginzburg, type III

$$\left(\left(\frac{P^2\left(\frac{ah(N-\Delta N)}{P^2}-\frac{1}{(N-\Delta N)}\right)}{a}\right)-\left(\frac{P^2\left(\frac{ahN}{P^2}-\frac{1}{N}\right)}{a}\right)\right)+PT=0$$

simplify

$$\frac{-ahN^2\Delta N + Nah\Delta N^2 - P^2\Delta N + PTN^2a - PTNa\Delta N}{(N-\Delta N)aN}=0$$

Isolate for ΔN

$$\Delta N = \frac{1}{2}\frac{1}{ahN}\left(ahN^2 + P^2 + PTNa - \left(\begin{array}{c}a^2h^2N^4 + 2ahN^2P^2 - 2a^2hN^3PT + P^4 + 2P^3TNa \\ + P^2T^2N^2a^2\end{array}\right)^{1/2}\right)$$

For factorizing the polynomial inside the square root the term $4TPha^2N^3$ is added:

$$Po = a^2h^2N^4 + 2ahN^2P^2 - 2a^2hN^3PT + P^4 + 2P^3TNa + P^2T^2N^2a^2 + 4TPha^2N^3$$

$$Po = a^2h^2N^4 + 2ahN^2P^2 + 2a^2hN^3PT + P^4 + 2P^3TNa + P^2T^2N^2a^2$$

$factor(Po)$;

$$\left(ahN^2 + P^2 + PTNa\right)^2$$

and

$$\Delta N = \frac{ahN^2 + P^2 + PTNa - \sqrt{(ahN^2 + P^2 + PTNa) - 4TPha^2N^3}}{2ahN}$$

Arditi-Akcakaya type III

$$\left(\left(\frac{P^m P^m \left(\frac{ah(N-\Delta N)}{P^m P^m} - \frac{1}{(N-\Delta N)}\right)}{a}\right) - \left(\frac{P^m P^m \left(\frac{ahN}{P^m P^m} - \frac{1}{N}\right)}{a}\right)\right) + PT = 0$$

simplify

$$\frac{-ahN^2 \Delta N + Nah\Delta N^2 - P^{2m}\Delta N + PTN^2 a - PTNa\Delta N}{(N-\Delta N)aN} = 0$$

Isolate for ΔN

$$\Delta N = \frac{1}{2}\frac{1}{ahN}\left(ahN^2 + P^{2m} + PTNa - \left(\begin{array}{c}a^2h^2N^4 + 2ahN^2 P^{2m} - 2a^2hN^3 PT + \left(P^{2m}\right)^2 + 2P^{2m} PTNa \\ + P^2T^2N^2a^2\end{array}\right)^{\frac{1}{2}}\right)$$

For factorizing the polynomial inside the square root the term $4TPha^2N^3$ is added:

$$Po = a^2h^2N^4 + 2ahN^2 P^{2m} - 2a^2hN^3 PT + \left(P^{2m}\right)^2 + 2P^{2m} PTNa + P^2T^2N^2a^2 + 4TPha^2N^3$$

$$Po = a^2h^2N^4 + 2ahN^2 P^{2m} + 2a^2hN^3 PT + \left(P^{2m}\right)^2 + 2P^{2m} PTNa + P^2T^2N^2a^2$$

$factor(Po);$

$$\left(ahN^2 + P^{2m} + PTNa\right)^2$$

and

$$\Delta N = \frac{ahN^2 + P^{2m} + PTNa - \sqrt{\left(ahN^2 + P^{2m} + PTNa\right)^2 - 4TPha^2N^3}}{2ahN}$$